\documentclass{elsarticle}
\usepackage{graphicx}
\usepackage{graphics}
\usepackage{epsfig}
\usepackage{amsmath}
\usepackage{amsfonts}
\usepackage{amssymb}
\usepackage{dcolumn}
\usepackage{bm}
\usepackage{longtable}
\usepackage{multirow}

\begin{document}

\title{Spin and orbital exchange interactions from Dynamical Mean Field Theory}

\author[RU]{A. Secchi \corref{cor1}}
\ead{a.secchi@science.ru.nl}
\ead{andrea.secchi@gmail.com}
\cortext[cor1]{Corresponding author}

\author[UH]{A. I. Lichtenstein}
\ead{alichten@physnet.uni-hamburg.de}

\author[RU]{M. I. Katsnelson}
\ead{m.katsnelson@science.ru.nl}

\address[RU]{Radboud University, Institute for Molecules and Materials, 6525 AJ Nijmegen, The Netherlands}
\address[UH]{Universitat Hamburg, Institut f\"ur Theoretische Physik, Jungiusstra{\ss}e 9, D-20355 Hamburg, Germany}

\date{\today}

\begin{abstract}

We derive a set of equations expressing the parameters of the magnetic interactions characterizing a strongly correlated electronic system in terms of single-electron Green's functions and self-energies. This allows to establish a mapping between the initial electronic system and a spin model including up to quadratic interactions between the effective spins, with a general interaction (exchange) tensor that accounts for anisotropic exchange, Dzyaloshinskii-Moriya interaction and other symmetric terms such as dipole-dipole interaction. We present the formulas in a format that can be used for computations via Dynamical Mean Field Theory algorithms.

\end{abstract}


\maketitle

\section{Introduction}

Describing a solid in terms of its magnetic properties requires the knowledge of an effective spin model which displays the same interesting physical properties as the many-electron Hamiltonian whose exact solution would give the complete description of the system. The determination of the form of the effective spin model and of the strength of the interactions between the constituent spins starting from the initial electronic model is, in general, a complicated many-body problem \cite{Auslender, Lichtenstein87, Stocks98, Katsnelson00, Katsnelson02, Bruno03, Udvardi03, Katsnelson04, Katsnelson10, Secchi13, Szilva13, Secchi15}.

We have recently derived expressions for the parameters of the magnetic interactions within an extended (multi-orbital) Hubbard model \cite{Secchi15}, in the presence of arbitrary relativistic couplings affecting the electronic degrees of freedom (such as spin-orbit, magnetic anisotropy, Zeeman coupling with an external magnetic field). The formulas presented in Ref.\cite{Secchi15}, after neglecting the vertices of two-electron Green's functions, are expressed in terms of single-electron (but fully interacting) Green's functions $G$ and the single-electron (hopping) Hamiltonian $T$. The use of the representation via $T$ \cite{Katsnelson10, Secchi15} for computations related to real materials requires the additional step of a tight-binding parametrization, which is implemented only in some methods of electronic structure calculations. On the other hand, a presentation of the formulas in terms of Green's functions $G$ and self-energies $\Sigma$ would make them more suitable for implementation via Dynamical Mean Field Theory (DMFT) \cite{Metzner89, Georges96, Kotliar06}, since \emph{any} DMFT calculation deals with $G$ and $\Sigma$. Writing the parameters in a way that explicitly exhibits self-energies, analogous to what was done in Refs.\cite{Katsnelson00, Katsnelson02, Secchi13}, also allows to explicitly include the approximation of local self-energy, which is the key assumption of DMFT. We here present the adaptation of the formulas for the exchange tensor to this scheme.

\section{Method and discussion}

We consider the extended multi-orbital Hubbard Hamiltonian \cite{Kanamori63, Hubbard65, Kugel82, Lichtenstein98, Georges13, Secchi15},
\begin{align}
\hat{H} = \sum_{o, \sigma, m} \sum_{o', \sigma', m'} \hat{\phi}^{\dagger}_{o, \sigma, m} T^{o, \sigma, m}_{o', \sigma', m'} \hat{\phi}^{o', \sigma', m'} + \hat{H}_V,
\label{Hamiltonian}
\end{align}
where the field operator $\hat{\phi}^{\dagger}_{o , \sigma , m }$ creates an electron with quantum numbers $\lbrace o, \sigma, m \rbrace$: $o$ refers to a set of the orbital indices (for a basis of localized Wannier wave functions, these are the atom index $a$, the principal atomic quantum number $n$ and the angular momentum quantum number $l$: $o \equiv \lbrace a, n, l \rbrace$), while $\sigma \in \lbrace \uparrow, \downarrow \rbrace$ and $m \in \lbrace -l , -l + 1, \ldots, l \rbrace$ are the third components of the intrinsic-spin and orbital angular momenta, respectively. Local angular momenta are measured with respect to local reference frames, which depend on $o$ and might not be collinear \cite{Secchi15}. The single-particle Hamiltonian matrix $T^{o, \sigma, m}_{o', \sigma', m'}$ is completely arbitrary, so it can include any relativistic single-electron terms (Zeeman coupling, spin-orbit, magnetic anisotropies). The interaction Hamiltonian $\hat{H}_V$ is assumed to be rotationally invariant \cite{Secchi15}.

The goal in Ref.\cite{Secchi15} was to map the model given by Eq.\eqref{Hamiltonian} onto an effective model of classical spins $\boldsymbol{e}_o$ including up to (arbitrary) quadratic interactions, with Hamiltonian
\begin{align}
H_{\mathrm{spin}} = \sum_o \boldsymbol{e}_o \cdot \boldsymbol{\mathcal{B}}_o            + \frac{1}{2} \sum_{o, o'} \sum_{\alpha, \alpha'} e_{o ,\alpha} e_{o', \alpha'} \mathcal{H}_{o o'}^{\alpha \alpha'} ,
\label{spin model}
\end{align}
determined by the exchange tensor $\mathcal{H}_{o o'}^{\alpha \alpha'} = \mathcal{H}_{o' o}^{\alpha' \alpha}$ (here and in the following $\alpha$ and $\alpha'$ are used to denote the space coordinates, e.g. $x, y, z$) and the effective magnetic field $\boldsymbol{\mathcal{B}}_o$. It is convenient to decompose the exchange tensor into the three vectors $\boldsymbol{\mathcal{J}}_{o o'} = \boldsymbol{\mathcal{J}}_{o' o}$ (anisotropic exchange), $\boldsymbol{\mathcal{D}}_{o o'} = - \boldsymbol{\mathcal{D}}_{o' o}$ (Dzyaloshinskii-Moriya interaction), and $\boldsymbol{\mathcal{C}}_{o o'} = \boldsymbol{\mathcal{C}}_{o' o}$ (symmetric non-diagonal exchange), defined as
\begin{align}
\mathcal{J}^{\alpha}_{o o'} \equiv  \mathcal{H}^{\alpha \alpha}_{o o'}, \quad  \mathcal{D}^{\alpha}_{o o'} \equiv \frac{1}{2} \sum_{\alpha' \alpha''} \varepsilon^{\alpha \alpha' \alpha''} \mathcal{H}^{\alpha' \alpha''}_{o o'}, \quad  \mathcal{C}^{\alpha}_{o o'} \equiv \frac{1}{2} \sum_{\alpha' \alpha''} \left| \varepsilon^{\alpha \alpha' \alpha''} \right| \mathcal{H}^{\alpha' \alpha''}_{o o'} ,
\end{align}
where $\varepsilon^{\alpha \alpha' \alpha''}$ is the completely anti-symmetric tensor of rank 3. The Heisenberg model is obtained as the particular case in which   $\mathcal{H}_{o o'}^{\alpha \alpha'} \equiv \delta^{\alpha \alpha'} \mathcal{J}_{o o'}$.

To perform the mapping, in Ref.\cite{Secchi15} we have derived the response of the thermodynamic potential of the electronic system under small spatially-dependent rotations of the spin quantization axes associated with each orbital spinor denoted by $o$, up to second order in the rotation angles. The derivation of such response involves path integration over the fermionic fields after the introduction of auxiliary bosonic degrees of freedom which express the amplitudes of rotations from an initial spin configuration; the coefficients of the interactions between the remaining bosons are put in correspondence with the parameters of the spin model \eqref{spin model} by imposing that the thermodynamic potential of the spin system after the spin rotations is equal to that of the electrons. Excluding the vertex contributions, the parameters of the spin model are expressed in terms of single-electron Green's functions (which of course include interaction effects) and the single-particle part of the electronic Hamiltonian, $T$.

This procedure is similar to the one previously adopted in Refs.\cite{Katsnelson00, Katsnelson02} for the case of quenched orbital moments, but in Ref.\cite{Secchi15} we have considered rotations of the local total spins $\hat{\boldsymbol{S}}_o = \hat{\boldsymbol{l}}_o + \hat{\boldsymbol{s}}_o$, where $\hat{\boldsymbol{l}}_o$ and $\hat{\boldsymbol{s}}_o$ are, respectively, the orbital and intrinsic angular momenta associated with the states $o$. More precisely, we have considered rotations in the space of the single-particle eigenfunctions of $\hat{\boldsymbol{S}}_o^2$ and $\hat{S}^z_o$, analogously to Ref.\cite{Katsnelson10}, while in Refs.\cite{Katsnelson00, Katsnelson02} the rotations affected the space of eigenfunctions of $\hat{\boldsymbol{s}}_o^2$ and $\hat{s}^z_o$. This allowed us to obtain formulas for the exchange tensor that can be separated into contributions coming from the interactions between spin-spin, orbital-orbital, or spin-orbital degrees of freedom of the electrons. It should be noted that this possibility is not applicable within Density Functional Theory (DFT) formulations, where observables are expressed in terms of the charge density and the intrinsic-spin density. The possibility of rotating local total spins is related to the representation of the electronic Hamiltonian in terms of localized wave functions, which implies a higher number of degrees of freedom with respect to DFT (related to the fact that the set of localized states would be over-complete in theory, or not even complete in practice due to truncation).

The computation of the magnetic parameters via DMFT is greatly simplified if they are formulated in terms of single-particle Green's functions and self-energies $\Sigma$ in magnetically ordered states, since this avoids the initial step of a tight-binding parameterization of the single-electron Hamiltonian $T$. To remove $T$ and introduce $\Sigma$, we use the equations of motion for Matsubara Green's functions (Dyson equations), which we write in general matrix notation as
\begin{align}
& \left( \omega - \mathrm{i} \mu \right) G(\mathrm{i} \omega) + \mathrm{i} T \cdot G(\mathrm{i} \omega) = 1  -   \Sigma(\mathrm{i} \omega) \cdot G(\mathrm{i} \omega) , \nonumber \\
& \left( \omega - \mathrm{i} \mu \right) G(\mathrm{i} \omega) + \mathrm{i}   G(\mathrm{i} \omega)  \cdot T  = 1 -  G(\mathrm{i} \omega) \cdot  \Sigma(\mathrm{i} \omega)   .
\label{Matsubara}
\end{align} 
These equations hold for the Matsubara Green's functions defined according to the following convention:
\begin{align}
G^1_2(\tau  ) \equiv - \mathrm{i} \left< \mathcal{T} \hat{\psi}^1(\tau) \, \hat{\psi}^{\dagger}_2  \right> \equiv \frac{1}{\beta} \sum_{\omega} G^1_2(\mathrm{i} \omega) \mathrm{e}^{- \mathrm{i} \omega   \tau  } ,
\end{align}
where $\omega = (2 n + 1) \pi / \beta$ is a fermionic Matsubara frequency. As a particular case, the single-electron density matrix is given by
\begin{align}
  \rho \equiv - \mathrm{i} \, G(\tau = 0^-) = -  \mathrm{i} \, \frac{1}{\beta}  \sum_{\omega} \mathrm{e}^{\mathrm{i} \omega 0^+} G(\mathrm{i} \omega)    .
\end{align}

We now have to distinguish between the magnetic parameters that can be computed from the second-order response in the rotation angles and those which are computed from the first-order response. From Ref.\cite{Secchi15}, we note that the former terms can all be written in terms of the following quantity:
\begin{align}
F^{\mathrm{XY}}_{o \alpha, o' \alpha'}  \equiv \, & - \delta_{ o o' } \frac{1}{2}  \mathrm{Tr}_{m, \sigma} \left( \left\{ S_{o \alpha}^{\mathrm{X}} ; S_{o \alpha'}^{\mathrm{Y}} \right\} \cdot \left\{ \rho ; T \right\}^o_o  \right)   \nonumber \\
& + \mathrm{Tr}_{m, \sigma} \left( S^{\mathrm{X}}_{o \alpha}  \cdot T^o_{o'} \cdot S^{\mathrm{Y}}_{o' \alpha'} \cdot \rho^{o'}_o  +   S^{\mathrm{Y}}_{o' \alpha'} \cdot T^{o'}_o \cdot S^{\mathrm{X}}_{o \alpha} \cdot \rho^o_{o'} \right) \nonumber \\
& +   \frac{1}{\beta}  \sum_{\omega} \mathrm{e}^{\mathrm{i} \omega 0^+} \mathrm{Tr}_{m, \sigma} \Big\{ S^{\mathrm{X}}_{o \alpha}  \cdot \left[ G(\mathrm{i} \omega ) \cdot T \right]^o_{o'} \cdot S^{\mathrm{Y}}_{o' \alpha'} \cdot \left[ G(\mathrm{i} \omega ) \cdot T \right]_o^{o'}  \nonumber \\
&   -   S^{\mathrm{X}}_{o \alpha}  \cdot   G(\mathrm{i} \omega )^o_{o'} \cdot S^{\mathrm{Y}}_{o' \alpha'} \cdot \left[ T \cdot G(\mathrm{i} \omega ) \cdot T \right]_o^{o'}   \nonumber \\
&   -   S^{\mathrm{X}}_{o \alpha}  \cdot  \left[ T \cdot G(\mathrm{i} \omega ) \cdot T \right]^o_{o'} \cdot S^{\mathrm{Y}}_{o' \alpha'} \cdot  G(\mathrm{i} \omega )_o^{o'}  \nonumber \\
&   + S^{\mathrm{X}}_{o \alpha}  \cdot \left[ T \cdot G(\mathrm{i} \omega )  \right]^o_{o'} \cdot S^{\mathrm{Y}}_{o' \alpha'} \cdot \left[ T \cdot G(\mathrm{i} \omega )  \right]_o^{o'}   \Big\} ,
\label{F}
\end{align}
where X, Y $\in \lbrace \mathrm{spin}, \mathrm{orb} \rbrace$ refer to either spin- or orbital- related terms, that is,
\begin{align}
S^{\mathrm{spin}}_{o \alpha} \equiv s_{o \alpha} \equiv \frac{1}{2} \sigma_{o \alpha}, \quad \quad  S^{\mathrm{orb}}_{o \alpha} \equiv l_{o \alpha},
\end{align}
where $s_{o \alpha}$ is an intrinsic spin matrix ($\sigma_{o \alpha}$ is a Pauli matrix), while $l_{o \alpha}$ is an orbital angular momentum matrix. In Eq.\eqref{F} we have used the notation $\left\{ A ; B \right\} \equiv A \cdot B + B \cdot A$ to denote the anti-commutator of the matrices $A$ and $B$; in the following we will also make use of $\left[ A ; B \right] \equiv A \cdot B - B \cdot A$ to denote the commutator.

From the Dyson equations \eqref{Matsubara}, we have (the frequency arguments of Green's functions $G$ and self-energies $\Sigma$ are implicit):
\begin{align}
&   T \cdot G  =  - \mathrm{i} \left[ 1 - \left( \omega - \mathrm{i} \mu \right) G  -  \Sigma  \cdot G  \right]   , \quad  G \cdot T  =  - \mathrm{i} \left[ 1 - \left( \omega - \mathrm{i} \mu \right) G  -  G  \cdot \Sigma  \right] ,  \nonumber \\
&   T \cdot G \cdot T   =  - \mathrm{i} T +  \Sigma  -    \Sigma   \cdot G  \cdot \Sigma   +      \left( \omega - \mathrm{i} \mu  \right)  \left( 1  -      \Sigma    \cdot  G  - G  \cdot \Sigma \right) - \left( \omega - \mathrm{i} \mu    \right)^2 G     , \nonumber \\
& \left[ T ; \rho \right] = \mathrm{Tr}_{\omega} \left[ \Sigma  ; G  \right] ,
\label{Matsubara 3}
\end{align} 
where we have introduced the notation
\begin{align}
\frac{1}{\beta}  \sum_{\omega} \mathrm{e}^{\mathrm{i} \omega 0^+}  f(\mathrm{i} \omega) \cdot g(\mathrm{i} \omega) \equiv \mathrm{Tr}_{\omega} (f \cdot g). 
\end{align}
Applying Eqs.\eqref{Matsubara 3} to Eq.\eqref{F}, we obtain
\begin{align}
   F^{\mathrm{XY}}_{o \alpha, o' \alpha'}  =   &  \frac{1}{2} \delta_{o o'}  \mathrm{Tr}_{\omega} \mathrm{Tr}_{m, \sigma} \Big(     \left\{ S^{\mathrm{X}}_{o \alpha}  ;    S^{\mathrm{Y}}_{o  \alpha'} \right\} \cdot \left\{     \Sigma  ; G    \right\}_o^{o}  \Big)
\nonumber \\
& - \mathrm{Tr}_{\omega} \mathrm{Tr}_{m, \sigma} \Big\{    S^{\mathrm{X}}_{o \alpha}  \cdot \left[        G  \cdot \Sigma  \right]^o_{o'} \cdot S^{\mathrm{Y}}_{o' \alpha'} \cdot \left[        G  \cdot \Sigma  \right]_o^{o'}  \nonumber \\
&   + S^{\mathrm{X}}_{o \alpha}  \cdot \left[      \Sigma  \cdot G  \right] ^o_{o'} \cdot S^{\mathrm{Y}}_{o' \alpha'} \cdot \left[        \Sigma  \cdot G  \right] _o^{o'}   
  -    S^{\mathrm{X}}_{o \alpha}  \cdot  \left[      \Sigma   \cdot G  \cdot \Sigma     \right]^o_{o'} \cdot S^{\mathrm{Y}}_{o' \alpha'} \cdot  G_o^{o'}   \nonumber \\  
&   -    S^{\mathrm{X}}_{o \alpha} \cdot  G^o_{o'}  \cdot S^{\mathrm{Y}}_{o' \alpha'}  \cdot  \left[      \Sigma   \cdot G  \cdot \Sigma     \right]_o^{o'}  \Big\}  \nonumber \\
& - \mathrm{Tr}_{\omega} \mathrm{Tr}_{m, \sigma} \Big(     S^{\mathrm{X}}_{o \alpha}  \cdot      \Sigma^o_{o'} \cdot S^{\mathrm{Y}}_{o' \alpha'} \cdot  G_o^{o'}     +   S^{\mathrm{Y}}_{o' \alpha'}  \cdot      \Sigma_o^{o'} \cdot S^{\mathrm{X}}_{o \alpha} \cdot  G^o_{o'} \Big)    .
\label{before DMFT}
\end{align}

We then consider the magnetic parameters determined from the first-order response. From Eqs.(68) of Ref.\cite{Secchi15}, we see that these are $\mathcal{B}_o^x$, $\mathcal{B}_o^y$, $\mathcal{D}_{o o'}^x$, $\mathcal{D}_{o o'}^y$, $\mathcal{C}_{o o'}^x$, and $\mathcal{C}_{o o'}^y$. The first-order response term (in the RHS of Eqs.(68) of Ref.\cite{Secchi15}) can be written as
\begin{align}
\mathcal{V}^{\mathrm{X}}_{o \alpha} & = \mathrm{i} \, \mathrm{Tr}_{m, \sigma} \left( S^{\mathrm{X}}_{o \alpha} \cdot \left[ \rho ; T \right]^o_o \right)  = \mathrm{i} \, \mathrm{Tr}_{m, \sigma} \mathrm{Tr}_{\omega} \left( S^{\mathrm{X}}_{o \alpha} \cdot \left[ G ; \Sigma \right]^o_o \right) \nonumber \\
& = \mathrm{i} \, \mathrm{Tr}_{m, \sigma} \mathrm{Tr}_{\omega} \sum_{o'} \left[ S^{\mathrm{X}}_{o \alpha} \cdot \left( G^o_{o'} \cdot \Sigma^{o'}_o - \Sigma^o_{o'} \cdot G^{o'}_o \right)  \right] .
\label{first}
\end{align}
By separating local and non-local terms, as well as taking into account the symmetries of the latter, it is then possible to identify the remaining parameters of the spin model. It should be noted that the parameters obtained with this procedure are \emph{not equivalent} to those expressed in terms of the single-electron Hamiltonian $T$ in Refs.\cite{Secchi15} and \cite{Katsnelson10}. These are different definitions, which respect the defining equations (68) of Ref.\cite{Secchi15}, but are more directly applicable for a DMFT implementation. 

In the next section we list the resulting formulas for the magnetic parameters.

\section{Results}

\subsection{Dzyaloshinskii-Moriya interaction}

The Dzyaloshinskii-Moriya parameters, given by the vector $\boldsymbol{\mathcal{D}}_{o o'}$, are written as  
\begin{align}
&            \left(  \mathcal{D}_{o o'}^{x}  \right)^{\mathrm{spin}}        =  \frac{ \mathrm{i} }{2 }       \mathrm{Tr}_{m, \sigma} \mathrm{Tr}_{\omega}  \Big[ s_{o x} \cdot \left( G^o_{o'} \cdot \Sigma^{o'}_o - \Sigma^o_{o'} \cdot G^{o'}_o    \right) \nonumber \\
& \quad \quad \quad \quad\quad  - s_{o' x} \cdot \left( G^{o'}_o \cdot \Sigma^o_{o'} -  \Sigma^{o'}_o \cdot G^o_{o'}     \right)      \Big]  ,    \nonumber \\
&            \left(  \mathcal{D}_{o o'}^{x}  \right)^{\mathrm{orb}}        =  \frac{ \mathrm{i} }{2 }       \mathrm{Tr}_{m, \sigma} \mathrm{Tr}_{\omega}  \Big[ l_{o x} \cdot \left( G^o_{o'} \cdot \Sigma^{o'}_o - \Sigma^o_{o'} \cdot G^{o'}_o    \right)  \nonumber \\
& \quad \quad \quad \quad\quad  - l_{o' x} \cdot \left( G^{o'}_o \cdot \Sigma^o_{o'} -  \Sigma^{o'}_o \cdot G^o_{o'}     \right)      \Big]  ,   
\end{align}
\begin{align}
&            \left(  \mathcal{D}_{o o'}^{y}  \right)^{\mathrm{spin}}        =  \frac{ \mathrm{i} }{2 }       \mathrm{Tr}_{m, \sigma} \mathrm{Tr}_{\omega}  \Big[ s_{o y} \cdot \left( G^o_{o'} \cdot \Sigma^{o'}_o - \Sigma^o_{o'} \cdot G^{o'}_o    \right) \nonumber \\
& \quad \quad \quad \quad\quad - s_{o' y} \cdot \left( G^{o'}_o \cdot \Sigma^o_{o'} -  \Sigma^{o'}_o \cdot G^o_{o'}     \right)     \Big]  ,    \nonumber \\
&            \left(  \mathcal{D}_{o o'}^{y}  \right)^{\mathrm{orb}}        =  \frac{ \mathrm{i} }{2 }       \mathrm{Tr}_{m, \sigma} \mathrm{Tr}_{\omega}  \Big[ l_{o y} \cdot \left( G^o_{o'} \cdot \Sigma^{o'}_o - \Sigma^o_{o'} \cdot G^{o'}_o    \right) \nonumber \\
& \quad \quad \quad \quad\quad - l_{o' y} \cdot \left( G^{o'}_o \cdot \Sigma^o_{o'} -  \Sigma^{o'}_o \cdot G^o_{o'}     \right)      \Big]  , 
\end{align}
\begin{align}
&   \left( \mathcal{D}_{o o'}^z \right)^{\mathrm{spin-spin}}     = \frac{1 }{2} \! \left(  F_{o x, o' y}^{\mathrm{spin}, \, \mathrm{spin}} - F_{o' x, o y}^{\mathrm{spin}, \, \mathrm{spin}} \right)    , \nonumber \\
&   \left( \mathcal{D}_{o o'}^z \right)^{\mathrm{orb-orb}}       = \frac{1 }{2} \! \left( F_{o x, o' y}^{\mathrm{orb}, \, \mathrm{orb}}  - F_{o' x, o y}^{\mathrm{orb}, \, \mathrm{orb}}  \right)   , \nonumber \\
&   \left( \mathcal{D}_{o o'}^z \right)^{\mathrm{spin-orb}}      = \frac{1 }{2} \! \left(  F_{o x, o' y}^{\mathrm{spin}, \, \mathrm{orb}}    +   F_{o x, o' y}^{\mathrm{orb}, \, \mathrm{spin}}  
 - F_{o' x, o y}^{\mathrm{spin}, \, \mathrm{orb}}    -   F_{o' x, o y}^{\mathrm{orb}, \, \mathrm{spin}} \right)       .
\end{align}

\subsection{Symmetric out-of-diagonal interactions}

The symmetric out-of-diagonal interaction parameters, given by the vector $\boldsymbol{\mathcal{C}}_{o o'}$ with $o \neq o'$, are written as  
\begin{align}
&            \left(  \mathcal{C}_{o o'}^{x}  \right)^{\mathrm{spin}}        =  \frac{ \mathrm{i} }{2 }       \mathrm{Tr}_{m, \sigma} \mathrm{Tr}_{\omega}  \Big[ s_{o x} \cdot \left( G^o_{o'} \cdot \Sigma^{o'}_o - \Sigma^o_{o'} \cdot G^{o'}_o    \right) \nonumber \\
& \quad \quad \quad \quad\quad  + s_{o' x} \cdot \left( G^{o'}_o \cdot \Sigma^o_{o'} -  \Sigma^{o'}_o \cdot G^o_{o'}     \right)      \Big]  ,    \nonumber \\
&            \left(  \mathcal{C}_{o o'}^{x}  \right)^{\mathrm{orb}}        =  \frac{ \mathrm{i} }{2 }       \mathrm{Tr}_{m, \sigma} \mathrm{Tr}_{\omega}  \Big[ l_{o x} \cdot \left( G^o_{o'} \cdot \Sigma^{o'}_o - \Sigma^o_{o'} \cdot G^{o'}_o    \right)  \nonumber \\
& \quad \quad \quad \quad\quad  + l_{o' x} \cdot \left( G^{o'}_o \cdot \Sigma^o_{o'} -  \Sigma^{o'}_o \cdot G^o_{o'}     \right)      \Big]  ,   
\end{align}
\begin{align}
&            \left(  \mathcal{C}_{o o'}^{y}  \right)^{\mathrm{spin}}        =  - \frac{ \mathrm{i} }{2 }       \mathrm{Tr}_{m, \sigma} \mathrm{Tr}_{\omega}  \Big[ s_{o y} \cdot \left( G^o_{o'} \cdot \Sigma^{o'}_o - \Sigma^o_{o'} \cdot G^{o'}_o    \right) \nonumber \\
& \quad \quad \quad \quad\quad + s_{o' y} \cdot \left( G^{o'}_o \cdot \Sigma^o_{o'} -  \Sigma^{o'}_o \cdot G^o_{o'}     \right)     \Big]  ,    \nonumber \\
&            \left(  \mathcal{C}_{o o'}^{y}  \right)^{\mathrm{orb}}        =  - \frac{ \mathrm{i} }{2 }       \mathrm{Tr}_{m, \sigma} \mathrm{Tr}_{\omega}  \Big[ l_{o y} \cdot \left( G^o_{o'} \cdot \Sigma^{o'}_o - \Sigma^o_{o'} \cdot G^{o'}_o    \right) \nonumber \\
& \quad \quad \quad \quad\quad + l_{o' y} \cdot \left( G^{o'}_o \cdot \Sigma^o_{o'} -  \Sigma^{o'}_o \cdot G^o_{o'}     \right)      \Big]  , 
\end{align}
\begin{align}
&   \left( \mathcal{C}_{o o'}^z \right)^{\mathrm{spin-spin}}     = - \frac{1 }{2}   \left(  F_{o x, o' y}^{\mathrm{spin}, \, \mathrm{spin}}  + F_{o' x, o y}^{\mathrm{spin}, \, \mathrm{spin}}  \right)      , \nonumber \\ 
&  \left( \mathcal{C}_{o o'}^z \right)^{\mathrm{orb-orb}}      = - \frac{1 }{2}  \left(  F_{o x, o' y}^{\mathrm{orb}, \, \mathrm{orb}}  + F_{o' x, o y}^{\mathrm{orb}, \, \mathrm{orb}} \right)    , \nonumber \\
&   \left( \mathcal{C}_{o o'}^z \right)^{\mathrm{spin-orb}}       = - \frac{1 }{2}   \left(   F_{o x, o' y}^{\mathrm{spin}, \, \mathrm{orb}}    + F_{o x, o' y}^{\mathrm{orb}, \, \mathrm{spin}} 
    +  F_{o' x, o y}^{\mathrm{spin}, \, \mathrm{orb}}    + F_{o' x, o y}^{\mathrm{orb}, \, \mathrm{spin}} \right)     .
\end{align}

\subsection{Local out-of-diagonal anisotropy}

The local out-of-diagonal anisotropy is given by the vector $\boldsymbol{\mathcal{C}}_{o o}$. We have
\begin{align}
&  \left( \mathcal{C}_{o o}^z \right)^{\mathrm{spin-spin}}      = - F_{o x, o y}^{\mathrm{spin,} \, \mathrm{spin}}  , \nonumber \\
&  \left( \mathcal{C}_{o o}^z \right)^{\mathrm{orb-orb}}        = - F_{o x, o y}^{\mathrm{orb,} \, \mathrm{orb}}   , \nonumber \\
&  \left( \mathcal{C}_{o o}^z \right)^{\mathrm{spin-orb}}     =    - F_{o x, o y}^{\mathrm{spin,} \, \mathrm{orb}} - F_{o x, o y}^{\mathrm{orb,} \, \mathrm{spin}}     ,
\end{align}
\begin{align}
& \left( \mathcal{C}_{oo}^x \right)^{\mathrm{spin}} = \mathrm{i} \, \mathrm{Tr}_{\omega} \mathrm{Tr}_{m, \sigma} \left( s_{o x} \cdot  \left[ G_o^o ; \Sigma_o^o \right] \right)  - \left( \mathcal{B}_o^y \right)^{\mathrm{spin}}  , \nonumber \\
& \left( \mathcal{C}_{oo}^x \right)^{\mathrm{orb}} = \mathrm{i} \, \mathrm{Tr}_{\omega} \mathrm{Tr}_{m, \sigma} \left( l_{o x} \cdot  \left[ G_o^o ; \Sigma^o_o\right] \right)  - \left( \mathcal{B}_o^y \right)^{\mathrm{orb}}  ,
\end{align}
\begin{align}
& \left( \mathcal{C}_{oo}^y \right)^{\mathrm{spin}} = - \mathrm{i} \, \mathrm{Tr}_{\omega} \mathrm{Tr}_{m, \sigma} \left( s_{o y} \cdot  \left[ G_o^o ; \Sigma^o_o\right] \right)  - \left( \mathcal{B}_o^x \right)^{\mathrm{spin}}  , \nonumber \\
& \left( \mathcal{C}_{oo}^y \right)^{\mathrm{orb}} = - \mathrm{i} \, \mathrm{Tr}_{\omega} \mathrm{Tr}_{m, \sigma} \left( l_{o y} \cdot  \left[ G_o^o ; \Sigma^o_o\right] \right)  - \left( \mathcal{B}_o^x \right)^{\mathrm{orb}}  ,
\end{align}
where the components of the effective magnetic field are
\begin{align}
& \boldsymbol{\mathcal{B}}_o^{\mathrm{spin}} \equiv g_{1/2}  \mu_{\mathrm{B}}  \boldsymbol{B}_o       \left[ \mathrm{Tr}_{ \sigma}  \left(  \boldsymbol{s}_{o } \mathrm{Tr}_{m } \rho^{o }_{o } \right)  \cdot \boldsymbol{u}^z_o \right]   , \nonumber \\
& \boldsymbol{\mathcal{B}}_o^{\mathrm{orb}} \equiv   g_l \mu_{\mathrm{B}}  \boldsymbol{B}_o      \left[ \mathrm{Tr}_{m}  \left(  \boldsymbol{l}_{o } \mathrm{Tr}_{ \sigma}  \rho^{o }_{o }  \right) \cdot \boldsymbol{u}^z_o   \right]  ,
\label{B sep}
\end{align}
with $g_{1/2}$ and $g_l$ being the intrinsic-spin and orbital $g$-factors, respectively, and $\boldsymbol{B}_o$ the value of the external magnetic field acting at the position of the orbitals $o$.

\subsection{Anisotropic exchange interactions}
\label{exch sep}

The anisotropic exchange parameters are given by the vector $\boldsymbol{\mathcal{J}}_{o o'}$ with $o \neq o'$. They are written as
\begin{align}
& \left(   \mathcal{J}^x_{o o'}  \right)^{\mathrm{spin-spin}}   =  F_{o y, o' y}^{\mathrm{spin}, \, \mathrm{spin}} , \nonumber \\
& \left(   \mathcal{J}^x_{o o'}  \right)^{\mathrm{orb-orb}}     =  F_{o y, o' y}^{\mathrm{orb}, \, \mathrm{orb}} , \nonumber \\
& \left(   \mathcal{J}^x_{o o'}  \right)^{\mathrm{spin-orb}}    =  F_{o y, o' y}^{\mathrm{spin}, \, \mathrm{orb}} + F_{o' y, o y}^{\mathrm{spin}, \, \mathrm{orb}} ,
\end{align}
\begin{align}
& \left(   \mathcal{J}^y_{o o'}  \right)^{\mathrm{spin-spin}}   =  F_{o x, o' x}^{\mathrm{spin}, \, \mathrm{spin}} , \nonumber \\
& \left(   \mathcal{J}^y_{o o'}  \right)^{\mathrm{orb-orb}}     =  F_{o x, o' x}^{\mathrm{orb}, \, \mathrm{orb}} , \nonumber \\
& \left(   \mathcal{J}^y_{o o'}  \right)^{\mathrm{spin-orb}}    =  F_{o x, o' x}^{\mathrm{spin}, \, \mathrm{orb}} + F_{o' x, o x}^{\mathrm{spin}, \, \mathrm{orb}}, 
\end{align}
and the terms related to $\mathcal{J}^{z}_{o o'}$ are given by the averages of the respective terms contributing to $\mathcal{J}^{x}_{o o'}$ and $\mathcal{J}^{y}_{o o'}$, since $\mathcal{J}^{z}_{o o'} = \left( \mathcal{J}^{x}_{o o'} + \mathcal{J}^{y}_{o o'} \right) / 2$.

It is instructive to consider the particular case where the self-energy is local not only in (atom) position space, but also diagonal in the principal quantum number ($n$) and orbital angular momentum ($l$) indices (we recall that $o \equiv \lbrace a, n, l \rbrace$), so that 
\begin{align}
\Sigma^o_{o'} \approx \delta^o_{o'} \Sigma_o ; 
\label{DMFT approx}
\end{align}
we obtain 
\begin{align}
&  \left(   \mathcal{J}^x_{o o'}  \right)^{\mathrm{spin-spin}}        \approx      - \mathrm{Tr}_{\omega} \mathrm{Tr}_{m, \sigma} \Big\{     \left[ s_{o y}  ;     \Sigma_o \right] \cdot G^o_{o'}    \cdot \left[ s_{o' y} ;   \Sigma_{o'} \right]  \cdot G_o^{o'}     \Big\}   , \nonumber \\
&  \left(  \mathcal{J}^{x}_{o o'} \right)^{\mathrm{orb-orb}}  \approx   - \mathrm{Tr}_{\omega} \mathrm{Tr}_{m, \sigma} \Big\{     \left[ l_{o y}  ;     \Sigma_o \right] \cdot G^o_{o'}    \cdot \left[ l_{o' y} ;   \Sigma_{o'} \right]  \cdot G_o^{o'}     \Big\} , \nonumber \\
 &  \left(  \mathcal{J}^{x}_{o o'} \right)^{\mathrm{spin-orb}}  
\approx  -     \mathrm{Tr}_{\omega} \mathrm{Tr}_{m, \sigma} \Big\{     \left[ s_{o y}  ;     \Sigma_o \right] \cdot G^o_{o'}    \cdot \left[ l_{o' y} ;   \Sigma_{o'} \right]  \cdot G_o^{o'}   \nonumber \\
& \quad \quad \quad \quad \quad \quad \quad   +   \left[ s_{o' y}  ;     \Sigma_{o'} \right] \cdot G_o^{o'}    \cdot \left[ l_{o y} ;   \Sigma_{o} \right]  \cdot G^o_{o'}    \Big\} ,
\end{align}
\begin{align}
&  \left(   \mathcal{J}^y_{o o'}  \right)^{\mathrm{spin-spin}}   \approx     - \mathrm{Tr}_{\omega} \mathrm{Tr}_{m, \sigma} \Big\{     \left[ s_{o x}  ;     \Sigma_o \right] \cdot G^o_{o'}    \cdot \left[ s_{o' x} ;   \Sigma_{o'} \right]  \cdot G_o^{o'} \Big\}  , \nonumber \\
&  \left(  \mathcal{J}^{y}_{o o'} \right)^{\mathrm{orb-orb}}    \approx     - \mathrm{Tr}_{\omega} \mathrm{Tr}_{m, \sigma} \Big\{     \left[ l_{o x}  ;     \Sigma_o \right] \cdot G^o_{o'}    \cdot \left[ l_{o' x} ;   \Sigma_{o'} \right]  \cdot G_o^{o'} \Big\} , \nonumber \\
 &    \left(  \mathcal{J}^{y}_{o o'} \right)^{\mathrm{spin-orb}}    \approx   -    \mathrm{Tr}_{\omega} \mathrm{Tr}_{m, \sigma} \Big\{     \left[ s_{o x}  ;     \Sigma_o \right] \cdot G^o_{o'}    \cdot \left[ l_{o' x} ;   \Sigma_{o'} \right]  \cdot G_o^{o'}  \nonumber \\
& \quad \quad \quad \quad \quad \quad \quad +  \left[ s_{o' x}  ;     \Sigma_{o'} \right] \cdot G_o^{o'}    \cdot \left[ l_{o x} ;   \Sigma_{o} \right]  \cdot G^o_{o'} \Big\} .
\end{align}
In this highly symmetric case, also assuming collinear magnetic states and that Green's functions and self-energies are diagonal in spin space, one obtains the isotropic exchange parameter 
\begin{align*} \left(   \mathcal{J}^x_{o o'}  \right)^{\mathrm{spin-spin}} = \left(   \mathcal{J}^y_{o o'}  \right)^{\mathrm{spin-spin}} = \left(   \mathcal{J}^z_{o o'}  \right)^{\mathrm{spin-spin}} \equiv \left(   \mathcal{J}_{o o'}  \right)^{\mathrm{spin-spin}} \end{align*} as
\begin{align}
\left(   \mathcal{J}_{o o'}  \right)^{\mathrm{spin-spin}}        \approx        \mathrm{Tr}_{\omega} \mathrm{Tr}_{m } \left(          \Sigma_o^{S} \cdot G^{o \uparrow}_{o' \uparrow}    \cdot     \Sigma^S_{o'}   \cdot G_{o \downarrow}^{o' \downarrow} 
+   \Sigma_o^{S} \cdot G^{o \downarrow}_{o' \downarrow}    \cdot     \Sigma^S_{o'}   \cdot G_{o \uparrow}^{o' \uparrow}    \right)   ,
\end{align}
where $\Sigma_o^{S} \equiv \left(\Sigma_o^{\uparrow} - \Sigma_o^{\downarrow} \right) / 2$, which is consistent with Eq.(109) of Ref.\cite{Secchi15} and with the previous literature \cite{Katsnelson00, Katsnelson02, Secchi13}.

\subsection{Local exchange interactions (diagonal anisotropy)}

The local exchange interaction parameters, or diagonal components of the local magnetic anisotropy, are given by $\boldsymbol{\mathcal{J}}_{o o}$. Within our rotational procedure, it is possible to determine only two of the three parameters as a function of the third one. Putting $\left( \alpha, \bar{\alpha} \right) = \left( x, y \right)$ or $\left( y, x \right)$, we obtain
\begin{align}
&   \left( \mathcal{J}_{o o}^{\bar{\alpha}} -  \mathcal{J}_{o o}^z \right)^{\mathrm{spin-spin}} = F_{o \alpha, o \alpha}^{\mathrm{spin,} \, \mathrm{spin}}   + \left( \mathcal{B}^{z}_{o} \right)^{\mathrm{spin}}   + \frac{1}{2} \sum_{o' \neq o} \left(    \mathcal{J}^{x}_{o o'} +  \mathcal{J}^{y}_{o o'}  , \right)^{\mathrm{spin-spin}}  , \nonumber \\
& \left( \mathcal{J}_{o o}^{\bar{\alpha}} -  \mathcal{J}_{o o}^z \right)^{\mathrm{orb-orb}} =  F_{o \alpha, o \alpha}^{\mathrm{orb,} \, \mathrm{orb}}    +  \left( \mathcal{B}^{z}_{o} \right)^{\mathrm{orb}}  + \frac{1}{2} \sum_{o' \neq o} \left(    \mathcal{J}^{x}_{o o'} +  \mathcal{J}^{y}_{o o'}  \right)^{\mathrm{orb-orb}} , \nonumber \\
& \left( \mathcal{J}_{o o}^{\bar{\alpha}} -  \mathcal{J}_{o o}^z \right)^{\mathrm{spin-orb}} =   F_{o \alpha, o \alpha}^{\mathrm{spin,} \, \mathrm{orb}} + F_{o \alpha, o \alpha}^{\mathrm{orb,} \, \mathrm{spin}}    + \frac{1}{2} \sum_{o' \neq o} \left(    \mathcal{J}^{x}_{o o'} +  \mathcal{J}^{y}_{o o'}  \right)^{\mathrm{spin-orb}} ,
\end{align}
where the non-local exchange terms are given in Section \ref{exch sep}, and the magnetic field components are defined in Eqs.\eqref{B sep}.

\section{Conclusions}

We have provided the formulas for the general exchange tensor expressing the quadratic magnetic interactions in strongly correlated systems, in a version that can be implemented via DMFT. The formulas allow to compute the effects due to intrinsic-spin and orbital degrees of freedom of the electrons on equal footing (the orbital magnetic moments are not quenched), and to distinguish between spin, orbital and spin-orbital interactions that contribute to the exchange tensor. The obtained formulas represent the extension to the relativistic case and the generalization to unquenched orbital magnetic moments of the well-known formulas for spin-only exchange interactions \cite{Katsnelson00}, which are recovered as a particular case. We remark that effects due to the non-locality of self-energies in position space are included in our theory both as presented in Ref.\cite{Secchi15} and as presented here; although they cannot be computed within DMFT, a possible approach to include them is via the Dual-Fermion scheme \cite{Rubtsov08}.

\section*{Acknowledgements}

This work is supported by the European Union Seventh Framework Programme under grant agreement No.281043 (FEMTOSPIN) and by the Deutsche Forschungsgemeinschaft under grant SFB-668.


\section*{References}

\begin{thebibliography}{30}





\bibitem{Auslender} M. I. Auslender and M. I. Katsnelson, Theor. Math. Phys. 51 (1982) 601; Solid State Commun. 44 (1982) 387.
\bibitem{Lichtenstein87} A. I. Liechtenstein, M. I. Katsnelson, V. P. Antropov, and V. A. Gubanov, J. Magn. Magn. Mater. 67 (1987) 65.
 
\bibitem{Stocks98} G. M. Stocks, B. Ujfalussy, X. Wang, D. M. C. Nicholson, W. A. Shelton, Y. Wang, A. Canning, and B. L. Gy\"orffy,  Philos. Mag. Part B 78 (1998) 665-673. 
\bibitem{Katsnelson00} M. I. Katsnelson and A. I. Lichtenstein, Phys. Rev. B 61 (2000) 8906.
\bibitem{Katsnelson02} M. I. Katsnelson and A. I. Lichtenstein, Eur. Phys. J. B: Condens. Matter 30 (2002) 9.
\bibitem{Bruno03} P. Bruno, Phys. Rev. Lett. 90 (2003) 087205.
\bibitem{Udvardi03} L. Udvardi, L. Szunyogh, K. Palot\'{a}s, and P. Weinberger, Phys. Rev. B 68 (2003) 104436.
\bibitem{Katsnelson04} M. I. Katsnelson and A. I. Lichtenstein, J. Phys.: Condens. Matter 16 (2004) 7439-7446.
\bibitem{Katsnelson10} M. I. Katsnelson, Y. O. Kvashnin, V. V. Mazurenko, and A. I. Lichtenstein, Phys. Rev. B 82 (2010) 100403(R) .
\bibitem{Secchi13} A. Secchi, S. Brener, A. I. Lichtenstein, and M. I. Katsnelson, Ann. Phys. 333 (2013) 221-271.
\bibitem{Szilva13} A. Szilva, M. Costa, A. Bergman, L. Szunyogh, L. Nordstr\"om, and O. Eriksson, Phys. Rev. Lett. 111 (2013) 127204.
\bibitem{Secchi15} A. Secchi, A. I. Lichtenstein, and M. I. Katsnelson, Ann. Phys. 360 (2015) 61-97.
 



\bibitem{Metzner89} W. Metzner and D. Vollhardt, Phys. Rev. Lett. 62 (1989) 324-327.
\bibitem{Georges96} A. Georges, G. Kotliar, W. Krauth, and M. J. Rozenberg, Rev. Mod. Phys. 68 (1996) 13.
\bibitem{Kotliar06} G. Kotliar, S. Y. Savrasov, K. Haule, V. S. Oudovenko, O. Parcollet, and C. A. Marianetti, Rev. Mod. Phys. 78 (2006) 865.



\bibitem{Kanamori63} J. Kanamori, Prog. Theor. Phys. 30 (1963) 275.
\bibitem{Hubbard65} J. Hubbard, Proc. Roy. Soc. A 285 (1965) 542.
\bibitem{Kugel82} K. I. Kugel and D. I. Khomskii, Sov. Phys. Uspekhi 25 (1982) 231. 
\bibitem{Lichtenstein98} A. I. Lichtenstein and M. I. Katsnelson, Phys. Rev. B 57 (1998) 6884.
\bibitem{Georges13} A. Georges, L. de' Medici, and J. Mravlje, Annu. Rev. Cond. Mat. Phys. 4 (2013) 137.

  


\bibitem{Rubtsov08} A. N. Rubtsov, M. I. Katsnelson, and A. I. Lichtenstein, Phys. Rev. B 77 (2008) 033101.



\end{thebibliography}
\end{document}